**Joint Inversion of DC Resistivity and MT Data using Multi-Objective Grey Wolf Optimization**

Rohan Sharma[*1], Divakar Vashisth[2], Kuldeep Sarkar[1] and Upendra Kumar Singh[1]

[1]*Department of Applied Geophysics, Indian Institute of Technology Dhanbad,* [2]*Department of Energy Science and Engineering, Stanford University*

email: * rohanlatha29@gmail.com

**Summary**

Joint inversion of geophysical datasets is instrumental in subsurface characterization and has garnered significant popularity, leveraging information from multiple geophysical methods. In this study, we implemented the joint inversion of DC resistivity with MT data using the Multi-Objective Grey Wolf Optimization (MOGWO) algorithm. As an extension of the widely-used Grey Wolf Optimization algorithm, MOGWO offers a suite of pareto optimal non-dominated solutions, eliminating the need for weighting parameters in the objective functions. This set of non-dominated predictions also facilitates the understanding of uncertainty in the predicted model parameters. Through a field case study in the region around Broken Hill in South Central Australia, the paper showcases MOGWO's capabilities in joint inversion, providing confident estimates of the model parameters (resistivity profiles), as indicated by a narrow spread in the suite of solutions. The obtained results are comparable to well established methodologies and highlight the efficacy of MOGWO as a reliable tool in geophysical exploration.

**Introduction**

Geophysical exploration plays a pivotal role in unraveling the mysteries beneath the Earth's surface. In the pursuit of understanding the subsurface, geophysical exploration methodologies have significantly evolved to leverage the synergies between different survey techniques. Several studies (Vozoff and Jupp, 1975; Singh *et al*., 2021; Liu *et al*., 2023) advocate for the integration of diverse geophysical methods to achieve better estimates of the subsurface properties. Among these, the joint inversion of Direct Current (DC) and Magnetotelluric (MT) data stands out as a sophisticated approach, offering a detailed understanding of the subsurface electrical and electromagnetic properties. The DC Resistivity method, widely applied in engineering and environmental geophysics (Sharma, 1997), and the MT method, recognized for its exploration potential in natural resource assessments (Simpson and Bahr, 2005), have individually contributed significantly to subsurface characterization. While DC surveys provide high-resolution information about near-surface electrical properties, MT surveys assist in delineating the electromagnetic properties of the deeper strata.

Multi-objective grey wolf optimization (MOGWO) extends the capabilities of the grey wolf optimization algorithm to solve joint inversion problems by generating a suite of non-dominated solutions without any need for weighting parameters for the objective functions (Vashisth *et al*., 2022). The spread in the suite of solutions helps in understanding the degree of uncertainty in the predicted model parameters, with a larger spread indicating higher uncertainty in the estimations and vice versa.

In this paper, we have performed the joint inversion of DC resistivity and MT data using MOGWO. We begin with a brief description of the physics required to simulate the DC and MT responses, followed by a discussion on the governing principles of MOGWO. Finally, we present a field case study of a region around Broken Hill in South Central Australia to demonstrate the efficacy of MOGWO.

**Theory**

Consider a 1D n-layered earth model. Let $\boldsymbol{\rho_n}$ and $\boldsymbol{h_n}$ denote the electrical resistivity and thickness of the $n^{th}$ layer. For DC resistivity sounding using the Schlumberger array, the apparent resistivity ($\rho_a$) is computed using equation (1) (Koefoed, 1979):

$$\rho_a(k) = k^2 \int_0^\infty \tau(\lambda)\lambda J_1(\lambda k)d\lambda \qquad (1)$$

where $k$ is half of the current electrode spacing (AB/2) and $J_1(\lambda k)$ is the first-order Bessel function. The resistivity transform $\tau(\lambda)$ is calculated using Pekeris recurrence relation (equation 2). For the first n-1 layers:

$$\tau_{n-1}(\lambda) = \frac{\tau_n(\lambda) + \boldsymbol{\rho_{n-1}} \tanh(\lambda \boldsymbol{h_{n-1}})}{1 + \tau_n(\lambda)\tanh(\lambda \boldsymbol{h_{n-1}})/\boldsymbol{\rho_{n-1}}} \qquad (2a)$$

For the $n^{th}$ layer:

$$\tau_n(\lambda) = \boldsymbol{\rho_n} \qquad (2b)$$

For MT sounding, the impedance ($Z_n$) is described as a function of frequency ($\omega$), using equation (3) (Sarkar *et al.*, 2023):

$$Z_n(\omega) = \frac{Z_{n+1} + T_n}{1 + S_n Z_{n+1}} \qquad (3)$$

The impedance is directly proportional to the true resistivity for the $n^{th}$ layer (equation 4).

$$Z_n = \omega\sqrt{\boldsymbol{\rho_n}} \qquad (4)$$

In equation (3), $T_n$ and $S_n$ can be written in terms of $\boldsymbol{\rho_n}$ and $\boldsymbol{h_n}$ using equation (5):

$$T_n = \omega\sqrt{\boldsymbol{\rho_n}}\tanh\left(\frac{\omega \boldsymbol{h_n}}{\sqrt{\boldsymbol{\rho_n}}}\right) \quad (5a); \qquad S_n = \frac{1}{\omega\sqrt{\boldsymbol{\rho_n}}}\left(\frac{\omega \boldsymbol{h_n}}{\sqrt{\boldsymbol{\rho_n}}}\right) \quad (5b)$$

The apparent resistivity ($\rho_a$) for MT sounding is determined using impedance ($Z$), magnetic permeability ($\mu$), and frequency ($\omega$) (equation 6):

$$\rho_a = \frac{1}{\mu\omega} Z^* Z \qquad (6)$$

The Grey Wolf Optimization (GWO) algorithm, introduced by Mirjalili *et al*. (2014), is a metaheuristic global optimization algorithm inspired by the social hierarchy and hunting behavior of grey wolves. It simulates the leadership and team dynamics of wolves in nature to find optimal solutions to complex problems, effectively balancing exploration and exploitation of the search space. Mirjalili *et al*. (2016) enhanced Grey Wolf Optimization (GWO) for multi-objective optimization by introducing two crucial steps. The first step involves maintaining an archive of non-dominated Pareto optimal solutions, updated through a specific scheme based on dominance criteria and a grid mechanism for managing crowded segments. The second step involves a leader selection strategy, selecting α, β, and δ solutions from the least crowded segments of the search space, aiding in identifying the best solutions for multi-objective problems. Vashisth *et al.* (2022) provided a detailed explanation of the Multi-Objective Grey Wolf Optimization (MOGWO) algorithm.

The MOGWO algorithm builds upon the foundation of GWO, maintaining its exploration and exploitation characteristics. However, MOGWO distinguishes itself by storing all non-dominated solutions in an archive, providing a suite of solutions instead of focusing solely on the three best solutions in each iteration, as in GWO. The suite of solutions generated by MOGWO also helps in understanding the uncertainty in the estimated model parameters, with a broader spread reflecting a higher degree of uncertainty. For the joint inversion of DC Resistivity and MT data, the L-2 norm of the difference between the observed data ($d_{DC}$ and $d_{MT}$) and the computed apparent resistivities ($A_{DC}\boldsymbol{m}$ and $A_{MT}\boldsymbol{m}$) constituted the objective function $f(\boldsymbol{m})$ as described in equation 7. $A_{DC}$ and $A_{MT}$ represent the forward modelling operators for DC resistivity and MT data, respectively. The vector $\boldsymbol{m}$ represents the model parameters (resistivity and thickness of each layer) that we want to estimate from joint inversion. Since both DC and MT apparent resistivities are dependent on the same model parameters ($m$), the mean of the non-dominated set of solutions could be regarded as a representative solution, to perform further quantitative analysis.

$$f(m) = [\,||d_{DC} - A_{DC}m||^2, ||d_{MT} - A_{MT}m||^2\,] \qquad (7)$$

**Case study: Broken Hill in South Central Australia**

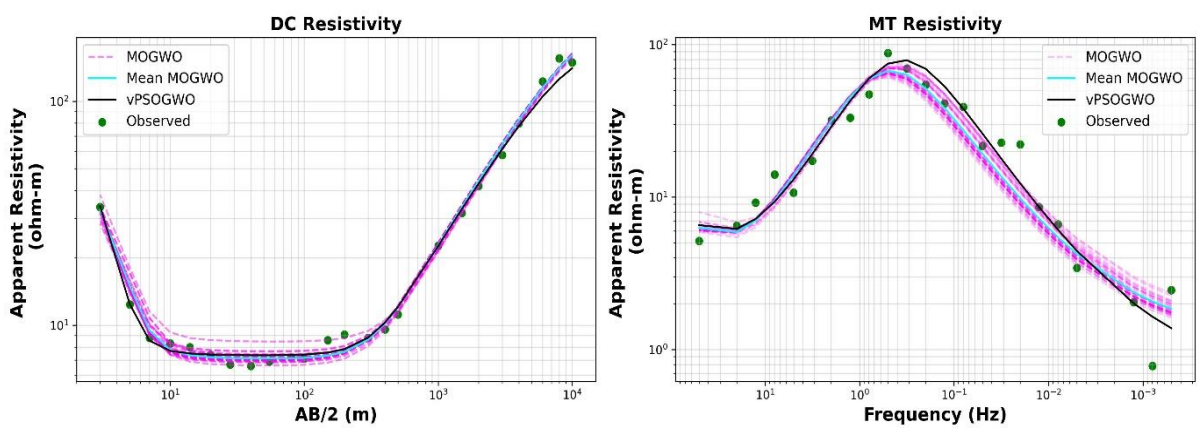

*Figure 1 Observed and computed DC and MT apparent resistivity curves by MOGWO and vPSOGWO. The observed apparent resistivity data is shown in green, and the apparent resistivities corresponding to the suite of MOGWO solutions are shown in purple. The blue and the black apparent resistivity curves correspond to the mean of the resistivity profiles predicted by MOGWO and vPSOGWO respectively.*

Joint inversion of DC and MT data was first implemented on the synthetic dataset used by Sarkar *et al.* (2023). The MOGWO predicted resistivities matched well with the true resistivity profile, with a slight increase in the spread at greater depths. Encouraged by these promising results, we also performed joint inversion of DC and MT datasets recorded from a location around Broken Hill in

South-Central Australia. The Schlumberger sounding was carried out with an electrode spacing of 20 km (Constable, 1985), followed by a comprehensive wide-band MT sounding (Cull, 1985) in the area. MOGWO was implemented using 23 grey wolves and the algorithm was executed for 500 iterations. The search space for model parameters (resistivity and thickness of each layer) was identical to the one employed in Sarkar *et al*. (2023). For the sake of comparison, the results of MOGWO were also compared with those obtained by the variable weight particle swarm optimizer - grey wolf optimizer (vPSOGWO) (Sarkar *et al*., 2023) and Occam's inversion (Constable *et al*., 1987).

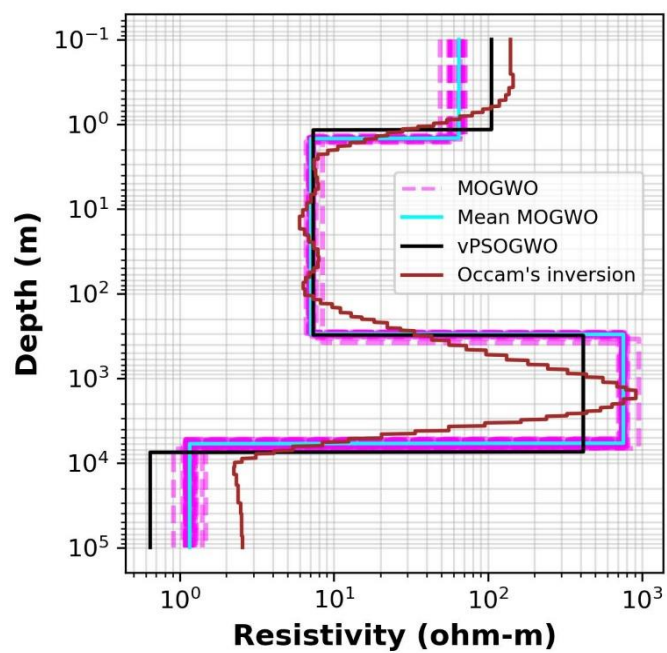

***Figure 2*** *Four-layered resistivity profiles resulting from the joint inversion of DC and MT data over a region near Broken Hill in South Central Australia. The suite of solutions obtained by MOGWO are represented by the purple curves and their corresponding mean resistivity profile is depicted in blue. Results obtained by vPSOGWO and Occam's inversion are indicated in black and brown respectively.*

## Results and Discussion

The joint inversion of DC and MT data using the MOGWO algorithm yielded 35 non-dominated pareto-optimal solutions in the archive. The MOGWO estimated resistivity profiles for the 4-layered Earth model, are demonstrated in Figure 2. The same figure also shows predictions from vPSOGWO and Occam's inversion. It can be seen that the suite of resistivity profiles predicted by MOGWO align closely with the results from vPSOGWO and Occam's inversion. The DC and MT apparent resistivity curves generated from the vPSOGWO and MOGWO predicted resistivities are shown in Figure 1. The DC and MT apparent resistivities corresponding to the suite of MOGWO predicted resistivity profiles matches well with the observed field data, indicating the successful implementation of joint inversion. The root mean square (RMS) errors between the observed and computed DC and MT apparent resistivity curves from the mean of the suite of MOGWO estimated resistivities, are 4.518 ohm-m and 7.369 ohm-m respectively. These errors are comparable to the ones calculated for vPSOGWO- 7.369 ohm-m for DC and 6.803 ohm-m for MT. Notably, MOGWO demonstrated a consistent and narrow spread throughout the suite of resistivity profiles, indicating a high degree of confidence in the estimated parameters. However, it is essential to acknowledge that the uncertainty estimate provided by MOGWO is qualitative, with plans for future work focusing on quantitative analysis of estimated uncertainties.

## Conclusions

In this study, we successfully demonstrated the capabilities of MOGWO by applying it to the joint inversion of DC resistivity and MT data acquired over a region around Broken Hill in South Central Australia. Our findings revealed results comparable to existing literature, affirming the efficacy of our approach. MOGWO for joint inversion offers a set of optimal, non-dominated solutions, with the spread of these solutions indicating the level of uncertainty in estimated model parameters. MOGWO can serve as an efficient and reliable tool for joint inversion applications in geophysical exploration.